\documentclass[sort&compress]{amsart}
\usepackage[utf8]{inputenc}
\usepackage{amsmath}
\usepackage{array, amsfonts}
\usepackage{amsthm}
\usepackage{amssymb}
\usepackage{enumerate}
\usepackage{lineno}
\usepackage[bookmarks=true]{hyperref}
\usepackage[foot]{amsaddr}
\usepackage{enumitem}
\usepackage{mathrsfs}
\usepackage{stackengine}
\usepackage{bookmark}
\usepackage{color,soul}
\usepackage[draft]{todonotes}
\usepackage{fancyhdr}
\usepackage{MnSymbol}
\usepackage{algorithm2e}
\usepackage[noend]{algpseudocode}
\usepackage{mathtools}
\usepackage[numbers]{natbib}

\definecolor{darkgreen}{rgb}{0,0.5,0}

\newcommand{\comment}[1]

\newcommand\myfunc[5]{%
	\begingroup
	\setlength\arraycolsep{0pt}
	#1\colon\begin{array}[t]{c >{{}}c<{{}} c}
		#2 & \to & #3 \\ #4 & \mapsto & #5 
	\end{array}%
	\endgroup}

\title{ \Large On the existence of hidden machines in computational time hierarchies}

\author{Felipe S. Abrah\~{a}o}
\address[Felipe S. Abrah\~{a}o, Klaus Wehmuth, Artur Ziviani]{National Laboratory for Scientific Computing (LNCC) \\ 25651-075 – Petropolis, RJ -- Brazil }
\email{fsa@lncc.br}

\author{Klaus Wehmuth}
\email{klaus@lncc.br}
\author{Artur Ziviani}
\email{ziviani@lncc.br}

\thanks{ Authors acknowledge the partial support from CNPq through their individual grants: F. S. Abrahão (313.043/2016-7), K. Wehmuth (312599/2016-1), and A. Ziviani (308.729/2015-3). Authors acknowledge the INCT in Data Science – INCT-CiD (CNPq 465.560/2014-8). Authors also acknowledge the partial support from FAPESP (2015/24493-1), and FAPERJ (E-26/203.046/2017).}

\fancypagestyle{style1}{
	\fancyhead{}
	\fancyhead[L]{ On the existence of hidden machines in computational time hierarchies }
	\fancyhead[R]{\thepage}
	\fancyfoot{}
}
\fancypagestyle{plain}{
	\fancyhead{}
	\fancyhead[C]{ \textit{Research Report} \\ National Laboratory for Scientific Computing (LNCC)  \\ \textbf{Extended version of the paper:} }
}
\newtheorem{lemma}{Lemma}[section]

\newtheorem{theorem}[lemma]{Theorem}

\theoremstyle{definition}
\newtheorem{definition}{Definition}[section]

\theoremstyle{remark}

\theoremstyle{definition}

\theoremstyle{definition}

\theoremstyle{remark}

\theoremstyle{remark}

\theoremstyle{theorem}

\begin{document}
	
\maketitle

\begin{abstract}\label{abstract}
	Challenging the standard notion of totality in computable functions, one has that, given any sufficiently expressive formal axiomatic system, there are total functions that, although computable and ``intuitively'' understood as being total, cannot be proved to be total. 
	In this article we show that this implies the existence of an infinite hierarchy of time complexity classes whose representative members are hidden from (or unknown by) the respective formal axiomatic systems.
	Although these classes contain total computable functions, there are some of these functions for which the formal axiomatic system cannot recognize as belonging to a time complexity class.
	This leads to incompleteness results regarding formalizations of computational complexity.
\end{abstract}



\newpage
\pagestyle{style1}

\section{Introduction}\label{sectionIntro}

The standard notion of a function being total is one of the defying counter-intuitive phenomena in axiomatizations of computer science (within Zermelo-Fraenkel with the Axiom of Choice (ZFC) or any other standard axiomatics for set theory).
As shown in \cite{Carnielli2008}, one of the damaging difficulties for axiomatizations based on first-order theories, encompassing Peano Arithmetic (PA) or ZFC, is the fact that, for every such a theory, there are total computable functions that are not recognizable as total recursive/computable
functions. 
This occurs in the context of fast-growing functions in such a way that, beyond a certain growth rate, some total computable functions cannot be proved to be total in sufficiently expressive formal axiomatic systems.
These functions can be computed for each input, but, although computable and total, the respective expression ``function $f$ is total'' cannot be a theorem. 
Thus, assuming the consistency of the chosen formal axiomatic systems, one can show there are incompleteness results regarding the totality of computable functions \cite{Carnielli2008}.

In this article, we show that the same kind of phenomenon also strikes the very foundation of computational complexity.
In particular, instead of formalizing the notion of ``function $f$ is total'', we investigate the notion of ``function $f$ belongs to a time complexity class'' in formal axiomatic systems.
Not only we demonstrate incompleteness of certain formulas about time complexity classes for axiomatizations based on first-order theories, but we also show the existence of a whole denumerable hierarchy of time complexity classes for which there are members that are not recognizable as belonging to a time complexity class.

\section{Preliminaries}

\subsection{Computable functions and formal axiomatic systems}\label{subsubsectionTMandAIT}

Regarding some basic notation, 
let $ \{ 0 , 1 \}^* $ be the set of all binary strings.
Let $ | x | $ denote the length of a string $ x \in  \{ 0 , 1 \}^* $.
Let $ (x)_2 $ denote the binary representation of the number $ x \in \mathbb{N} $. 
In addition, let $ (x)_{L} $ denote the representation of the number $ x \in \mathbb{N} $ in language $ L $.
%

%
%
%
%
%
%

%

\begin{definition}\label{BdefFunctionU}
	Let $ \mathbf{M}(x) $ denote the output of a Turing machine (TM) $\mathbf{M}$ when $x$ is given as input in its tape. 
	Thus, $ \mathbf{M}(x) $ denotes a \emph{partial recursive} function
	\[
	\myfunc{ \varphi_{\mathbf{M}} }{ \mathbf{L} }{ \mathbf{L} }{ x }{ y = \varphi_{\mathbf{M}}(x) } \text{ ,}
	\]  
	\noindent where $\mathbf{L}$ is a language.

\end{definition}

In particular, $ \varphi_{\mathbf{U}}(x) $ is the \emph{universal} partial recursive function \cite{Rogers1987} and $ \mathbf{L_U} $ denotes a universal programming language for a universal Turing machine $\mathbf{U}$.
Note that, if $x$ is a non-halting program on $\mathbf{M}$, then this function $\mathbf{M}(x)$ is undefined for $x$.
Wherever $ \, n \in \mathbb{N} $ or $ n \in \{ 0 , 1 \}^* $ appears in the domain or in the codomain of a partial (or total) recursive function
\[
\myfunc{ \varphi_{ \mathbf{M} } }{ \mathbf{L} }{ \mathbf{L} }{ x }{ y = \varphi_{ \mathbf{M} }(x) } \text{ ,}
\]
\noindent where $ \mathbf{M} $ is a Turing machine, running on language $\mathbf{L}$, it actually denotes $ \left( n \right)_{ \mathbf{L} } $.

Let $ \{ e \} $  denote the partial computable function $ \varphi_{\mathbf{M}} $ for which $ e $ is its index (e.g., its G\"{o}del number) such that the Kleene's predicate $ T( e , x , z ) $ has a well defined value $z$ whenever there is a $y$ such that $ \varphi_{ \mathbf{M} }( x ) \equiv \mathbf{M}( x ) \equiv \{ e \}(x) = y $.
Note that we are employing $ e $ for symbolizing the encoding of a \emph{deterministic} TM, for example by employing G\"{o}del numbers.

With respect to weak asymptotic dominance of function $f$ by a function $g$, 
we employ the usual $f(x)=\mathbf{O}( g(x) )$ for the big \textbf{O} notation when $f$ is asymptotically upper bounded by $g$; 
and with respect to strong asymptotic dominance by a function $g$, we employ the usual $f(x)=\mathbf{o}( g(x) )$ when $g$ dominates $f$. 

Let $ S $ denote any sufficiently expressive formal axiomatic system (FAS) in the language $ L $ such that there is an interpretation of Zermelo-Fraenkel with the Axiom of Choice (ZFC) into $ S $.
Note that there is then an interpretation of Peano Arithmetic (PA) into $S$, and therefore Kleene's predicate is definable in the language of $S$.
As in \cite{Carnielli2008}, we denote the well-formed formula (wff) expressing ``$ \{ e \} $ is a total function'' in the language of $S$ (i.e., $ \forall x \exists z  T( e , x , z )  $) by $ [ \{ e \} \text{ is total} ] $.
From the Kleene's predicate $ T( e , x , z ) $, one can also construct a wff (e.g., denoted by $ [ \{ e \}(x) = y ] $) in the language of $ S $ that defines the function relation $ \{ e \}(x) = y $ wherever there is a $y$ given a $x$.
And, since $ S $ encompasses  PA, then $ \left[ \mathrm{Prov}_S\left( h , x \right) \right] \in L $, where $ \mathrm{Prov}_S\left( h , x \right) $ is the wff in PA that represents the existence of a proper deductive proof of the wff encoded by $ x $ in which $ h $ is the G\"{o}del number of the sequence proof steps from the axioms of $ S $ that end in the wff encoded by $x$ \cite{Abrahao2011anat,Smorynski1977}.
     
In addition, we have that the partial computable function $ \textit{\textbf{time}}( e , x ) $ that returns the computation running time (or \emph{running time} for short) of the partial computable function  $ \{ e \} $, if it defined for $ x $, can also be defined in the language of $S$, which we denote by $ [ \textit{\textbf{time}}( e ,  x  ) = y ] $ \cite{Abrahao2016nat,Abrahao2015ams}.
Note that $ \textit{\textbf{time}}( e , x ) $ is the running time of deterministic TMs with input $x$ and that, since $ \{ e \} $ is computable, 
\begin{center}
	$ \textit{\textbf{time}}( e , x ) $ is defined \textit{iff} $ \{ e \}(x) $ is also defined.
\end{center}
For example, a TM of index $ t_e $ that computes $ \textit{\textbf{time}}( e ,  x  ) $ can be the one that receives $ x $ as input, emulates $ \mathbf{M} $ of index $ e $ with input $ x $, with an additional tape for counting the number of computation steps of $ e $, and then returns the value from this additional counting tape whenever $ \mathbf{M} $ reaches a halting state.
In particular, we know that the running time for calculating this emulation (wherever the value of $ \{ e \}(x) $ is defined) is in the worst case cubic, that is,
$ \textit{\textbf{time}}( t_e ,  x  ) = \mathbf{O}\left( \left( \textit{\textbf{time}}( e ,  x  ) \right)^3 \right)  $ \cite{Papadimitriou1994}.

\subsection{Computational time complexity}

We base our notion of time complexity classes in a traditional manner as in the literature \cite{Papadimitriou1994,Rich2007}.
The time complexity class $ \mathbf{TIME}\left( f( | x | ) \right) $ is the class of decision problems that can be solved by deterministic TMs in $\mathbf{O}( f( | x | ) ) $ computation steps.
Analogously, we also have the time complexity class $ \mathbf{NTIME}\left( f( | x | ) \right) $ for non-deterministic TMs.
This way, the polynomial time complexity $ \textbf{P}\textbf{-TIME} $ of all decision problems that can be computed by deterministic TMs in polynomial computation steps as a function of the input length is given by the parametrized time complexity $ \textbf{P}\textbf{-TIME} = \bigcup_{ j > 0 } \mathbf{TIME}\left( | x |^j  \right)  $ and the same applies analogously to the non-deterministic case $ \textbf{NP}\textbf{-TIME} $.

In addition to decision problems (i.e.,  TMs that always decide correctly whether an input belongs or not to a language $L$ ), one also has \emph{function problems}.
Thus, the class $ \textbf{FP}\textbf{-TIME} $ is the class of partial functions $ \{ e \} $ such that, for every $x$ for which there is $ y = \{ e \}(x) $, one has that $ \textit{\textbf{time}}( e ,  x  ) = \mathbf{O}\left( | x |^k  \right)$ is dominated by a polynomial, where $ k \in \mathbb{N} $.
More specifically, the time complexity class $ \mathbf{FTIME}\left( f( | x | ) \right) $ is the class of function problems that can be solved by deterministic TMs in $\mathbf{O}( f( | x | ) ) $ computation steps.
Analogously, the class $ \textbf{FNP}\textbf{-TIME} $ is the class of partial relations $ R $ such that, for every $x$ for which there is $y$ with $ ( x , y ) \in R $, one has that there is a non-deterministic polynomially time-bounded TM that can find at least one $y$ such that $ ( x , y ) \in R $.

The class of all function problems in $ \textbf{FNP}\textbf{-TIME} $ that are total (i.e., for every $x$ there is at least one $y$ with $ ( x , y ) \in R \in \textbf{FNP}\textbf{-TIME} $) is denoted by $ \textbf{TFNP}\textbf{-TIME} $.
The same way, the class of all function problems in $ \textbf{FP}\textbf{-TIME} $ that are total (i.e., for every $x$ there is \emph{only one} $y$ with $ ( x , y ) \in f \in \textbf{FP}\textbf{-TIME} $) is denoted by $ \textbf{TFP}\textbf{-TIME} $.
And, as usual, the time complexity class $ \mathbf{TFTIME}\left( f( | x | ) \right) $ is the class of total function problems that can be solved by deterministic TMs in $\mathbf{O}( f( | x | ) ) $ computation steps.
An interesting theorem that already is known to hold is that  \cite{Papadimitriou1994}:
\[ 
\textbf{FP}\textbf{-TIME} = \textbf{FNP}\textbf{-TIME} \iff \textbf{P}\textbf{-TIME} = \textbf{NP}\textbf{-TIME} 
\text{ .}
\]

It is straightforward to show that, since both the time complexity of deciding whether a TM halts or not in polynomial time and the time complexity of emulating a TM's running time are cubic, one has that
\[  \textbf{TFP}\textbf{-TIME} \subseteq \textbf{FP}\textbf{-TIME} \subseteq \textbf{TFNP}\textbf{-TIME} \subseteq \textbf{FNP}\textbf{-TIME} 
\]
hols.

All of these classes can also be analogously extended to the exponential time complexity classes $ \textbf{EXP}\textbf{-TIME} $, $ \textbf{FEXP}\textbf{-TIME} $, $ \textbf{TFEXP}\textbf{-TIME} $, and so on.
In this article, we will deal only with total function problems.
In particular, our proofs hold for deterministic Turing machines.

\section{A logic dependence of computer science from the totality of functions}

In \cite{Carnielli2008}, it is shown the existence of function that defies the axiomatization of computer science, so that even ``fairly intuitive'' notions cannot be grasped by sufficiently expressive formal axiomatic systems. 
A computable function can be constructively defined, i.e., programmed on an universal TM, such that there is a diagonalization procedure that eventually lands on every total computable function $ f $ that $ S $ can prove it is total (i.e., $ S \vdash  [ f \text{ is total} ] $) and then maximizes its value.
In other words, it is a computable function that eventually grows faster than any other total computable function that $ S $ proves is total.
A function $ F $ with these properties was constructed as follows:

\begin{definition}
	Let $ F $ be a function whose values are given by the TM $ \mathbf{M}_F $ that receives $ n \in \mathbb{N} $ as input and:
	\begin{enumerate}
		\item enumerates all $e$ such that $ S \vdash \left[ \mathrm{Prov}_S\left( y , [ \{ e \} \text{ is total} ] \right) \right] $ and $ y \leq n $;
		
		\item constructs a finite list $ \left( e_1 , \dots , e_z \right) $ of these functions $ \{ e \} $, where $ z \in \mathbb{N} $;
		
		\item returns $ \max\left\{ y \; \vert y = \{ e \}(x) + 1 \land x \leq n \right\} $.
		
	\end{enumerate}
\end{definition}

Note that we have that $ F $ is intuitively total in the sense that, if $ S $ is indeed consistent, then every computable function indexed by $e$ that $ S $ proves is total will always return a value. 
This way, a maximization from the constructed list of these $e$'s will be always possible.

Moreover, $F$ is clearly a partial computable function.
There is a TM $ \mathbf{M}_F $ such that, for every given particular $ n $ as input, it will effectively calculate the value of $ F(n) $.
The question is whether or not the fact that $F$ is total can be proved in $ S $:
and the answer is negative.
To this end, just note that, if $ S $ eventually proves $  [ \{ e_F \} \text{ is total} ] $, where $ e_F $ is the index of the TM $ \mathbf{M}_F  $, there will be a $ n_0 $ from which, for every $ n \geq n_0 $, $ F(n) > F( n ) $. 
Thus, one can individually check that, for each natural number $ n $, $ F(n) $ exists and can be computed. 
However, $ S $  cannot “join” all those results together to show that $ F $ is total.

This kind of diagonalization on growth rate can be also found in the Busy Beaver function.
Specially in the form 
$ BB : \, \mathbb{N} \to \, \mathbb{N} $ as a function that returns the largest integer that a program $ p \in \mathbf{L_U}  $ with length $ \leq N \in \mathbb{N} $ can output running on machine $ \mathbf{U} $ \cite{Abrahao2017publishednat}.
Such a Busy Beaver function has several interesting properties.
For example, although function $ BB $ eventually grows faster than any other computable function $ f_c $ (that is, for every computable function $ f_c   : \mathbb{N} \to \mathbb{N}$, there is $ N_0 \in \mathbb{N} $ such that, for every $ N \geq N_0 $, $ BB(N) > f_c(N) $), it is a \emph{scalable} uncomputable function, i.e., for every  $ N \in \mathbb{N} $, there is a program $ p \in \mathbf{L_U} $ such that $ \mathbf{U}( p ) = BB(N) $ (in particular, $ \left| p \right| \leq N $).
Moreover, function $ BB $ is an incompressible function, i.e., there is constant $ c \in \mathbb{N} $ such that, for every $ N \in \mathbb{N} $, $ \mathbf{I}_A( BB(N) ) \geq N - c $, where $ \mathbf{I}_A( \cdot ) $ is the algorithmic information or algorithmic complexity of an object \cite{Chaitin2004,Abrahao2018dextendedarxiv2020reportnat}.

Now, note that $BB$ eventually grows faster than function $F$, since $F$ is computable.
Indeed, function $ F $ defined in \cite{Carnielli2008} also has that ``ungraspable but eventually reachable'' property that $BB$ has.
In other words, although function $ F $ eventually grows faster than any other $S$-provenly total computable function $ f_c $, it is a \emph{scalable} total computable function.
However, what one may deem to even more counter-intuitive in the case of $F$, is that it can be indeed computed by a effectively constructible TM, wheres $BB$ is an uncomputable function.
Thus, as $BB$ works like a ``ceiling'' function for every computable function---function $F$ included---, function $F$ also works like a ``ceiling'' function, but for $S$-provenly total computable functions.

This way, as pointed out in \cite{Carnielli2008}, any attempt to find a FAS for computer science in which the naive intuition of e.g. totality of functions can be grasped, faces difficulties of the same order of incompleteness in mathematical logic.
Other incompleteness phenomena related to axiomatization of computer science were also presented in \cite{Carnielli2008}, such as the relation between the totality of function $F$ and $\Sigma_1$-soundness and recognition of sets of polynomially time-bounded TMs.
With respect to the latter, we shall show in this article other incompleteness phenomena in computer science that some may deem to be even more dramatic.
In other words, going even further into computer science, we shall show later on that this ``odd'' function $F$ is also related to incompleteness phenomena in one of the central subjects in theoretical computer science: computational complexity (or algorithm analysis).

%
%
%

\section{General time complexity classes}

We aim to show that there are time complexity classes of function problems that contain functions associated with so fast-increasing running time that the expression in $S$ that defines ``this function belongs to a time complexity class'' cannot be proved, although it would be intuitively true in a standard model of arithmetic for example.
For this purpose, the main idea is that a recognizable time complexity class in $ S $ is the one for which every TM $ M $ in this class can be proved to belong to by $ S $.
In this article, we are only dealing with deterministic TMs and with total function problems.
Thus, more formally:

\begin{definition}\label{defGeneraltimecomplexityclassFT}
	Let $ L $ be the language of the FAS $ S $.
	Let  $ \left[ \{ e \} \in  \textbf{TF}X\textbf{-TIME} \right] $ denote 
	\begin{equation*}
		\begin{aligned}
			\exists z ( z \in X \land \forall x \exists h , k (  [ \textit{\textbf{time}}( e , x ) = k ] \land k \leq h \land  [ \{ z \}( | x | ) = h ] ) ) 
		\end{aligned}
	\end{equation*}
	\noindent in the language $ L $, where $ X $ is a free variable.
	We say that $ S $ \emph{recognizes} $ \{ e \} $ as belonging to a deterministic time complexity class $ X $ iff $ S \vdash \left[ \{ e \} \in  \textbf{TF}X\textbf{-TIME} \right] $.
\end{definition}

In general, $X$ can be any set definable in the language of $S$.
This is the case of for example $ X = \{ f_p \} $, $ X = \{ f_{exp} \} $, or $ X = \{ f \} $, where $ f_p $ is a polynomial, $ f_{exp} $ is an exponentiation, and $ f $ is any time-constructible \cite{Rich2007} (or proper complexity function \cite{Papadimitriou1994}).
Thus, we immediately obtain from Definition~\ref{defGeneraltimecomplexityclassFT} that 
\[ \textbf{TF}\{ f \}\textbf{-TIME} = \textbf{TFTIME}( f ) \text{ .} \]
One can also define $X$ as a union of functions in order to cover what is called \emph{parametrized time complexity classes}.
For example: 
if $ X = \{ f \vert f(x) = \mathbf{O}\left( x^k \right) \land x,k \in \mathbb{N} \} $ is the set of all polynomials, then $ \textbf{TF}X\textbf{-TIME} = \textbf{TFP}\textbf{-TIME} $;
if $ X = \{ f \vert f(x) = \mathbf{O}\left( k^{ \left( x^m \right) } \right) \land x,k,m \in \mathbb{N} \} $ is the set of all exponentiation, then $ \textbf{TF}X\textbf{-TIME} = \textbf{TFEXP}\textbf{-TIME} $; and so on.

Now, instead of a specific set $X$ of running time functions, one could also investigate whether or not $S$ can recognize a TM belonging to an arbitrary deterministic time complexity class.
Indeed, this is easily defined by:

\begin{definition}\label{defGeneralbelongingtoanexistingtimecomplexityclass}
	Let $ L $ be the language of the FAS $ S $.
	We say that $ S $ \emph{recognizes} $ \{ e \} $ as belonging to \emph{at least one} deterministic time complexity class iff $ S \vdash \exists X \left[ \{ e \} \in  \textbf{TF}X\textbf{-TIME} \right] $.
\end{definition}

\section{Non-recognizable Turing machines in a time complexity hierarchy}

In this section, we tackle the main objective in this article.
We aim to investigate total time-bounded TMs for which $ S $ cannot recognize as belonging to \emph{at least one} deterministic time complexity class.
Indeed, TM $ \mathbf{M}_F $ is one of these:

\begin{theorem}\label{thmCentral1}
	There is a total computable function $ \{ e \} $ such that, if $S$ is consistent, then  
	\begin{equation}\label{equationMainnonprovability}
	S \nvdash \exists X \left[ \{ e \} \in  \mathbf{TF}X\mathbf{-TIME} \right]
	\end{equation}
	and
	\begin{equation}\label{equationTrivialnonprovability}
	S \nvdash \neg \exists X \left[ \{ e \} \in  \mathbf{TF}X\mathbf{-TIME} \right] \text{ .}
	\end{equation}
	
	\begin{proof}
		We employ in this proof the total computable function $F$.
		Let $ \{ e \} $ denote $ F $.
		First, Equation~\eqref{equationTrivialnonprovability} trivially follows from the fact that, if 
		$ S \vdash \neg \exists X \left[ \{ e \} \in  \textbf{TF}X\textbf{-TIME} \right] $ and $ S $ is consistent, then we would have that $F$ is not a total function in an standard model of $ S $ for example.
		Then, it remains to prove that $ S \nvdash \exists X \left[ \{ e \} \in  \textbf{TF}X\textbf{-TIME} \right] $.
		To this end, suppose $ S \vdash \exists X \left[ \{ e \} \in  \textbf{TF}X\textbf{-TIME} \right] $ holds.
		Now, note from Definitions~\ref{defGeneralbelongingtoanexistingtimecomplexityclass} and \ref{defGeneraltimecomplexityclassFT} that
		\begin{equation}
			S \vdash \left( \exists X \left[ \{ e \} \in  \textbf{TF}X\textbf{-TIME} \right] \to [ \textit{\textbf{time}}( e , \cdot ) \text{ is total} ] \right)
		\end{equation}
		and
		\begin{equation}
			S \vdash \left( [ \textit{\textbf{time}}( e , \cdot ) \text{ is total} ] \to [ \{ e \} \text{ is total} ] \right)
		\end{equation} 
		hold.
		Therefore, we would have that
		\begin{equation}
			S \vdash  [ \{ e \} \text{ is total} ] 
					\text{ ,}
		\end{equation} 
		which we already know it is false. 
	\end{proof}
	
\end{theorem}

In addition, we can extend function $F$ in order to include $ \exists X \left[ F \in  \textbf{TF}X\textbf{-TIME} \right] $:

\begin{definition}
	Let $ F^{(1)} $ be a function whose values are given by the TM $ \mathbf{M}_{ F^{(1)} } $ that receives $ n \in \mathbb{N} $ as input and:
	\begin{enumerate}
		\item enumerates all $e$ such that $ S \vdash \left[ \mathrm{Prov}_{ S^{(1)} }\left( y , [ \{ e \} \text{ is total} ]  \right) \right] $ and $ y \leq n $, where $ S^{(1)} = S + \exists X \left[ F \in  \textbf{TF}X\textbf{-TIME} \right] $;
		
		\item constructs a finite list $ \left( e_1 , \dots , e_z \right) $ of these functions $ \{ e \} $, for some $ z \in \mathbb{N} $;
		
		\item returns $ \max\left\{ y \; \vert y = \{ e \}(x) + 1 \land x \leq n \right\} $.
		
	\end{enumerate}
\end{definition}

And, by continuing this process, we will obtain a sequence $ F , F^{(1)}, F^{(2)}, \dots $ of functions defined respectively for $ S, S^{(1)}, S^{(2)}, \dots $, where $ S = S^{(0)} $, $ F = F^{(0)} $ and
\begin{equation*}
S^{(k+1)} = S^{(k)} + \exists X \left[ F^{(k)} \in  \textbf{TF}X\textbf{-TIME} \right]\text{ .}
\end{equation*}
This way, for each iteration one obtains another $ \exists X \left[ F^{(k+1)} \in  \textbf{TF}X\textbf{-TIME} \right] $ such that 
\begin{equation}\label{equationMainnonprovabilitygeneral}
S^{(k+1)} \nvdash \exists X \left[ F^{(k+1)} \in  \textbf{TF}X\textbf{-TIME} \right]
\end{equation}
and
\begin{equation}\label{equationTrivialnonprovabilitygeneral}
S^{(k+1)} \nvdash \neg \exists X \left[ F^{(k+1)} \in  \textbf{TF}X\textbf{-TIME} \right] \text{ .}
\end{equation}

Furthermore, since the time complexity overhead in simulating a TM is cubic, we will have that
\begin{equation}\label{}
S^{(k+1)} \vdash \exists X \left[ \textit{\textbf{time}}( e_{ F^{(k)} } , \cdot ) \in  \textbf{TF}X\textbf{-TIME} \right]
\end{equation} 
In fact, any time-constructible total function $ f $ composed with $ \textit{\textbf{time}}( e_{ F^{(k+1)} } , \cdot ) $ can be proved by $ S^{(k)} $ to be in a time complexity class, that is
\begin{equation}\label{}
S^{(k+1)} \vdash \exists X \left[ f\left( \textit{\textbf{time}}( e_{ F^{(k)} } \right) , \cdot ) \in  \textbf{TF}X\textbf{-TIME} \right]
\text{ .}
\end{equation} 
For example, one has that
\begin{equation}\label{equationTimecomplexityofhyperexpcomposed}
S^{(k+1)} \vdash \exists X \left[ \text{hyperexp}\left( \textit{\textbf{time}}( e_{ F^{(k)} } \right) , \cdot ) \in  \textbf{TF}X\textbf{-TIME} \right]
\text{ ,}
\end{equation} 
where $ \text{hyperexp}( \cdot ) $ is the hyperexponentiation function.

Thus, the following theorem establishes a infinite denumerable time hierarchy built from $ F , F^{(1)}, F^{(2)}, \dots $:

\begin{theorem}\label{thmCentral2}
	Let $ X $ be an arbitrary set of functions such that $ \mathbf{O}\left( \textit{\textbf{time}}( e_{ F^{(k)} } , \cdot ) \right) $ is the faster growing function in $ X $.
	Then, there is a set $ Y $ of functions such that $ X \subseteq Y $ and
	$ F^{(k+1)} \nin \textbf{TF}X\textbf{-TIME} $ and $ F^{(k+1)} \in \textbf{TF}Y\textbf{-TIME} $.
	
	\begin{proof}
		The main idea of the proof is to show that, if $ F^{(k+1)} \in \textbf{TF}X\textbf{-TIME} $, then it would eventually grow faster than it could.
		Suppose that $ F^{(k+1)} \in \textbf{TF}X\textbf{-TIME} $.
		Then, by our construction of $ X $, we would have $ \textit{\textbf{time}}( e_{ F^{(k+1)} } , x ) = \mathbf{O}\left( \textit{\textbf{time}}( e_{ F^{(k)} } , x ) \right) $.
		But, from Equation~\eqref{equationTimecomplexityofhyperexpcomposed}, we have that 
		\begin{equation*}\label{}
			S^{(k+1)} \vdash \exists X \left[ \text{hyperexp}\left( \textit{\textbf{time}}( e_{ F^{(k)} } \right) , \cdot ) \in  \textbf{TF}X\textbf{-TIME} \right]
			\text{ .}
		\end{equation*}
		Now, remember that $ F^{(k+1)} $ eventually grows faster than any $ S^{(k+1)} $-provenly total computable function.
		Therefore, the value returned by $ F^{(k+1)} $ would eventually become so large that even the length of its binary representation (which is in a logarithmic order of the value) would strongly dominate the number of computation steps (which by construction is in $ \mathbf{O}\left( \textit{\textbf{time}}( e_{ F^{(k)} } , x ) \right) $) that are generating it, which is a contradiction.
		Finally, in order to prove there is a $Y$ such that $ F^{(k+1)} \in \textbf{TF}Y\textbf{-TIME} $ and $ X \subseteq Y $, we define $ Y \coloneqq X \bigcup \{ \mathbf{O}\left( \textit{\textbf{time}}( e_{ F^{(k+1)} } , \cdot ) \right) \} $.
	\end{proof}
\end{theorem}

\section{Conclusion}

By formalizing the general notion of a computable function belonging to a time complexity class in first-order language, we showed in Theorem~\ref{thmCentral1} that, for any sufficiently expressive formal axiomatic system (e.g., ZFC) believed to be consistent, there is a function that belongs to a time complexity class but that this formal axiomatic system cannot prove it belongs to a time complexity class.
Such a phenomenon is a new type of incompleteness, but that occurs in computational complexity expressed by formal axiomatic systems.
In addition, we showed in Theorem~\ref{thmCentral2} that there is an infinite sequence of total computable functions, each within respectively higher time complexity classes, such that none of these functions can be recognized as belonging to a time complexity class by sufficiently expressive formal axiomatic systems.
In other words, for every sufficiently expressive formal axiomatic system believed to be consistent, there are total computable functions that are hidden from belonging to time complexity classes, which in turn are ordered in an infinite denumerable time complexity hierarchy.
Thus, together with the logic dependence of computer science with respect to the concept of totality as in \cite{Carnielli2008}, the present article highlights the presence of such a logic dependence also in computational complexity.

\small 
\bibliographystyle{plainnat}
\bibliography{2.2.1-CompleteRefs-Felipe}

\end{document}